\let\ifarxiv=\iftrue     
\pdfoutput=1

\ifarxiv

\documentclass[12pt,a4paper]{article}

\usepackage{amsmath,amssymb}

\ifnum\pdfoutput=1
\usepackage[bookmarks=true,hyperfigures=true]{hyperref}
\else

\fi

\usepackage[nosort]{cite}
\usepackage[bulletsep]{collect}

\ifx\hypersetup\sadfkjashdfkxja\else
\hypersetup{pdftitle={The SU(2|3) Undynamic Spin Chain}}
\hypersetup{pdfauthor={Niklas Beisert}}
\hypersetup{pdfsubject={}}
\hypersetup{pdfkeywords={AdS/CFT, Integrable Spin Chain, Dynamic Spin Chain}}
\hypersetup{plainpages=false}
\hypersetup{bookmarksnumbered=true}
\hypersetup{pdfstartview=FitH}
\hypersetup{pdfpagemode=UseNone}
\hypersetup{colorlinks=false}
\hypersetup{citebordercolor={.5 1 .5}}
\hypersetup{urlbordercolor={.5 1 1}}
\hypersetup{linkbordercolor={1 .7 .7}}
\fi

\makeatletter
\let\old@makecaption=\@makecaption
\def\@makecaption{\small\old@makecaption}
\makeatother

\else

\documentclass[seceq]{ptptex}

\def\paragraph{\subsection}

\fi

\makeatletter
\let\old@startsection=\@startsection
\renewcommand{\@startsection}[6]{\old@startsection{#1}{#2}{#3}{#4}{#5}{#6\mathversion{bold}}}
\makeatother

\allowdisplaybreaks[3]

\numberwithin{equation}{section}

\let\oldPhi=\Phi
\let\oldPsi=\Psi
\let\oldGamma=\Gamma
\let\oldDelta=\Delta
\let\oldSigma=\Sigma
\let\oldLambda=\Lambda
\let\oldTheta=\Theta
\let\oldPi=\Pi
\renewcommand{\Phi}{\mathnormal{\oldPhi}}
\renewcommand{\Psi}{\mathnormal{\oldPsi}}
\renewcommand{\Gamma}{\mathnormal{\oldGamma}}
\renewcommand{\Sigma}{\mathnormal{\oldSigma}}
\renewcommand{\Delta}{\mathnormal{\oldDelta}}
\renewcommand{\Theta}{\mathnormal{\oldTheta}}
\renewcommand{\Lambda}{\mathnormal{\oldLambda}}
\renewcommand{\Pi}{\mathnormal{\oldPi}}

\newcommand{\gen}[1]{\mathfrak{#1}}
\newcommand{\mdl}[1]{\mathcal{#1}}
\newcommand{\bgspin}{\mathcal{Z}}

\newcommand{\order}[1]{\mathcal{O}(#1)}

\newcommand{\Integers}{\mathbb{Z}}

\newcommand{\Reals}{\mathbb{R}}

\ifx\genfrac\sdflkaj

\else

\fi
\newcommand{\sfrac}[2]{{\textstyle\frac{#1}{#2}}}
\newcommand{\half}{\sfrac{1}{2}}

\newcommand{\indup}[1]{_{\mathrm{#1}}}

\newcommand{\lrbrk}[1]{\left(#1\right)}
\newcommand{\bigbrk}[1]{\bigl(#1\bigr)}

\newcommand{\comm}[2]{[#1,#2]}
\newcommand{\lrcomm}[2]{\left[#1,#2\right]}
\newcommand{\acomm}[2]{\{#1,#2\}}

\newcommand{\bigeval}[1]{#1\big|}
\newcommand{\eval}[1]{#1|}
\newcommand{\lreval}[1]{\left.#1\right|}
\newcommand{\PTerm}[2]{\big\{{\textstyle\genfrac{}{}{0pt}{}{#1}{#2}}\big\}}
\newcommand{\PYTerm}[4]{\big\{{\textstyle\genfrac{}{}{0pt}{}{#1}{#2}\big|\genfrac{}{}{0pt}{}{#3}{#4}}\big\}}

\newcommand{\vspan}[1]{\langle #1\rangle}
\newcommand{\bigvspan}[1]{\bigl\langle #1\bigr\rangle}
\newcommand{\state}[1]{\mathopen{|}#1\mathclose{\rangle}}

\newcommand{\alg}[1]{\mathfrak{#1}}

\newcommand{\nln}{\nonumber\\}
\newcommand{\nl}[1][0pt]{\nonumber\\[#1]&\hspace{-4\arraycolsep}&\mathord{}}

\newcommand{\earel}[1]{\mathrel{}&\hspace{-2\arraycolsep}#1\hspace{-2\arraycolsep}&\mathrel{}}
\newcommand{\eq}{\earel{=}}
\newcommand{\beq}{\begin{equation}}
\newcommand{\eeq}{\end{equation}}

\def\[{\begin{equation}}
\def\]{\end{equation}}
\def\<{\begingroup\ifarxiv\else\arraycolsep1pt\fi\begin{eqnarray}}
\def\>{\end{eqnarray}\endgroup\ignorespaces}

\makeatletter
\def\mr@ignsp#1 {\ifx\:#1\@empty\else #1\expandafter\mr@ignsp\fi}%
\newcommand{\multiref}[1]{\begingroup
\xdef\mr@no@sparg{\expandafter\mr@ignsp#1 \: }%
\def\mr@comma{}%
\@for\mr@refs:=\mr@no@sparg\do{\mr@comma\def\mr@comma{,}\ref{\mr@refs}}%
\endgroup}
\makeatother

\newcommand{\hypref}[2]{\ifx\href\asklfhas #2\else\href{#1}{#2}\fi}

\renewcommand{\eqref}[1]{(\multiref{#1})}

\ifx\href\asklfhas\newcommand{\href}[2]{#2}\fi
\newcommand{\arxivlink}[1]{\href{http://arxiv.org/abs/#1}{arxiv:#1}}

\begin{document}

\ifarxiv

\begin{flushright}\footnotesize
\texttt{\arxivlink{0807.0099}}\\
\texttt{AEI-2008-042}%
\end{flushright}
\vspace{1cm}

\begin{center}%
{\Large\textbf{\mathversion{bold}%
The $\mathfrak{su}(2|3)$ Undynamic Spin Chain}\par}
\vspace{1cm}%

\textsc{Niklas Beisert}\vspace{5mm}%

\textit{Max-Planck-Institut f\"ur Gravitationsphysik\\%
Albert-Einstein-Institut\\%
Am M\"uhlenberg 1, 14476 Potsdam, Germany}\vspace{3mm}%

\texttt{nbeisert@aei.mpg.de}
\par\vspace{1cm}

\textbf{Abstract}\vspace{7mm}

\begin{minipage}{12.7cm}
The integrable spin chain found in 
perturbative planar $\mathcal{N}=4$ supersymmetric gauge theory is dynamic. 
Here we propose a reformulation which removes the dynamic effects
in order to make the model more accessible 
to an algebraic treatment.
\end{minipage}

\end{center}

\vspace{1cm}
\hrule height 0.75pt
\vspace{1cm}


\else

\markboth{Niklas Beisert}{The su(2$/$3) Undynamic Spin Chain}

\title{The $\mathfrak{su}(2|3)$ Undynamic Spin Chain}


\author{Niklas \textsc{Beisert}}

\inst{Max-Planck-Institut f\"ur Gravitationsphysik\\%
Albert-Einstein-Institut\\%
Am M\"uhlenberg 1, 14476 Potsdam, Germany}



\abst{%
The integrable spin chain found 
in perturbative planar $\mathcal{N}=4$ supersymmetric 
gauge theory is dynamic. Here we propose a
reformulation which removes the dynamic effects
in order to make the model more accessible 
to an algebraic treatment.
}

\maketitle

\fi

\section{Introduction}

The study of integrable structures in planar perturbative 
$\mathcal{N}=4$ supersymmetric Yang--Mills theory 
following the works \cite{Lipatov:1997vu,Minahan:2002ve,Beisert:2003yb}
has led to the discovery of an exciting integrable spin chain model.
It displays several unusual and novel features with respect to 
the established integrable spin chains: 
First of all, the spin chain is perturbatively long-ranged \cite{Beisert:2003tq}. 
In other words, the Hamiltonian 
not only acts on nearest-neighbouring spins, 
but also on longer blocks of adjacent spins.
The range is controlled by the perturbative order
in a coupling constant $g\approx 0$.
Moreover the chain is dynamic \cite{Beisert:2003ys},
that is, the Hamiltonian 
may dynamically change the number of spin sites of the chain. 
Finally, the Hamiltonian is an inseparable part of
the symmetry algebra. Consequently, 
all the above features of the Hamiltonian apply to the 
symmetry generators as well.
In addition it can be remarked that the
spin module is non-compact and graded
into bosons and fermions. 

Despite these complications, it
appears that the Hamiltonian is completely integrable
\cite{Lipatov:1997vu,Minahan:2002ve,Beisert:2003yb,Beisert:2003tq,Beisert:2003ys,Serban:2004jf}.
Because it is homogeneous and acts locally, 
one can apply the asymptotic coordinate Bethe ansatz
\cite{Sutherland:1978aa,Staudacher:2004tk}.
The form of the asymptotic Bethe equations \cite{Beisert:2005fw}
is fully constrained by symmetry considerations \cite{Beisert:2007ty},
merely one phase function remains undetermined.
Imposing a further crossing symmetry \cite{Janik:2006dc,Arutyunov:2006iu}
together with inspiration from the dual 
superstring theory on $AdS_5\times S^5$ \cite{Maldacena:1998re}
and its integrable structure \cite{Bena:2003wd}
one arrives at a viable proposal for the phase
\cite{Beisert:2006ib,Beisert:2006ez}
which has since passed several
highly non-trivial tests
\cite{Bern:2006ew,Benna:2006nd,Basso:2007wd,Roiban:2007dq}.

Note well that the above mentioned asymptotic Bethe equations
describe the spectrum only up to certain finite-size corrections, 
see \cite{Fiamberti:2007rj,Fiamberti:2008sh}
and references therein, 
yet to be understood from the integrable model point of view.
A conceivable path towards the exact finite-size spectrum
is to fully understand the algebraic structure 
underlying the integrable spin chain model. 
One of the obstacles are the dynamic effects 
for which the conventional algebraic structures 
appear to be inapplicable.

In this note we consider the prototypical
dynamic sector of the $\mathcal{N}=4$ SYM spin chain
with $\alg{su}(2|3)$ symmetry \cite{Beisert:2003ys}.%
\footnote{The $\mathcal{N}=6$ superconformal Chern--Simons theory \cite{Aharony:2008ug}
with $\alg{osp}(6|4,\Reals)$ symmetry 
has an analogous $\alg{su}(2|3)$ sector \cite{Minahan:2008hf}.
The results of \cite{Beisert:2003ys} 
and of this note are general and they
also apply to this model 
with some minor modifications regarding, e.g.\ the coupling constant
and the embedding.} 
We shall propose an undynamic reformulation 
where length fluctuations are absent for a 
large part of the algebra including the Hamiltonian.
This is meant to facilitate an eventual algebraic treatment
of the model.
We will start with a review
of the $\alg{su}(2|3)$ sector,
then propose the undynamic reformulation
and finally discuss the implications
and potential pitfalls.

\section{Dynamic Chain}

Let us start by reviewing 
the (apparently) integrable $\alg{su}(2|3)$ dynamic spin chain
constructed in \cite{Beisert:2003ys}.

\paragraph{Hilbert Space.}

The spin at each site can be in three bosonic states
$\state{\phi^a}$ with $a=1,2,3$, and two fermionic states
$\state{\psi^\alpha}$ with $\alpha=1,2$.
Thus the graded spin module $\mathcal{V}$ is spanned 
by the five states
\[
\mdl{V}=
\bigvspan{\phi^1,\phi^2,\phi^3\mathpunct{\big|}\psi^1,\psi^2}.
\]
The Hilbert space $\mdl{H}$ of the spin chain model 
is given by the direct sum of cyclic chain spaces $\mdl{H}_L$
of arbitrary length $L$
\[\label{eq:Hilbert}
\mdl{H}=\bigoplus_{L=1}^\infty \mdl{H}_L,\qquad
\mdl{H}_L=\bigeval{\mdl{V}^{\otimes L}}\indup{cyclic}.
\]
The space $\eval{\mdl{V}^{\otimes L}}\indup{cyclic}$ 
represents the subspace of $\mdl{V}^{\otimes L}$ 
on which the graded cyclic shift operator acts 
as the identity.
The dynamic nature of the model consists in the fact that
the Hamiltonian (as well as the other symmetry generators)
acts as an endomorphism of $\mdl{H}$ and not of the individual $\mdl{H}_L$'s,
in other words, the length of the spin chain is a dynamic quantity.
Furthermore our spin chain is homogeneous
which entails the restriction to cyclic states: 
Homogeneous operators commute with the graded permutation
whose spectrum $\exp(2\pi i\Integers/L)$
crucially depends on the length. The only common eigenvalue
on chains of $L$ and $L+1$ is $1$ and thus dynamic homogeneous 
models must be based on cyclic states.

\paragraph{Symmetry Algebra.}

The symmetry of the dynamic chain is assumed to be $\alg{su}(2|3)$.
This algebra is spanned by 
the $\alg{su}(3)$ generators $\gen{R}^a{}_b$ ($\gen{R}^a{}_a=0$),
the $\alg{su}(2)$ generators $\gen{L}^a{}_b$ ($\gen{L}^a{}_a=0$),
the fermionic generators $\gen{Q}^\alpha{}_b$ and $\gen{S}^a{}_\beta$
and finally the Hamiltonian $\gen{H}$.
The Lie superalgebra is given by the canonical Lie brackets
for $\alg{su}(3)$ and $\alg{su}(2)$ and the supercharges 
transform in (anti)fundamental representations, e.g.\
\[
\comm{\gen{R}^a{}_b}{\gen{Q}^\gamma{}_d}=
-\delta^a_d\gen{Q}^\gamma{}_b
+\sfrac{1}{3}\delta^a_b\gen{Q}^\gamma{}_d.
\]
The non-trivial brackets among the supercharges are given by 
\[
\acomm{\gen{Q}^\alpha{}_b}{\gen{S}^c{}_\delta}=
\delta^\alpha_\delta\gen{R}^c{}_b
+\delta^c_b\gen{L}^\alpha{}_\delta
+\delta^\alpha_\delta\delta^c_b\gen{H}.
\]
Finally, the weights of the supercharges with respect
to the Hamiltonian read
\[\label{eq:QSeng}
\comm{\gen{H}}{\gen{Q}^\alpha{}_b}=+\sfrac{1}{6} \gen{Q}^\alpha{}_b,
\qquad
\comm{\gen{H}}{\gen{S}^a{}_\beta}=-\sfrac{1}{6} \gen{S}^a{}_\beta.
\]
%

\paragraph{Representation.}

We want to construct a family of representations $\gen{J}(g)$ of
$\alg{su}(2|3)$ on the Hilbert space $\mdl{H}$
parametrised by a coupling constant $g$.
The coupling constant $g$ is assumed to be small
and we shall treat the representation 
as a perturbation series around $g=0$
\[
\gen{J}(g)=\gen{J}_0+g \gen{J}_1+g^2 \gen{J}_2+\ldots
\]
At leading order the representation $\gen{J}_0$ 
is given by the standard tensor product 
of fundamental representations of $\alg{su}(2|3)$
\[
\begin{array}{rcl}
(\gen{R}_0)^a{}_b\eq
\PTerm{a}{b}-\sfrac{1}{3}\delta^a_b\PTerm{c}{c},
\\[3pt]
(\gen{L}_0)^\alpha{}_\beta\eq
\PTerm{\alpha}{\beta}-\sfrac{1}{2}\delta^\alpha_\beta\PTerm{\gamma}{\gamma},
\end{array}\quad
\begin{array}{rcl}
(\gen{Q}_0)^\alpha{}_b\eq
\PTerm{\alpha}{b},
\\[3pt]
(\gen{S}_0)^a{}_\beta\eq
\PTerm{a}{\beta},
\end{array}\quad
\gen{H}_0=
\sfrac{1}{3}\PTerm{a}{a}
+\sfrac{1}{2}\PTerm{\alpha}{\alpha}.
\]
The interaction symbols $\PTerm{\cdot}{\cdot}$ have the following meaning: 
For example, $\PTerm{\beta}{a}$ picks any boson $\phi^a$ 
from the chain and replaces it by a fermion $\psi^\beta$. 
Here Latin and Greek indices 
refer to bosons and fermions, respectively. 
A homogeneous sum over all sites
with proper grading is implicit in this notation.

The $\alg{su}(3)$ and $\alg{su}(2)$ representations 
are finite-dimensional and cannot be deformed continuously
\[
\gen{R}^a{}_b(g)=(\gen{R}_0)^a{}_b,\qquad
\gen{L}^\alpha{}_\beta(g)=(\gen{L}_0)^\alpha{}_\beta.
\]
The representation of supercharges is deformed at all order in $g$,
the first correction reads
\[\label{eq:QS1}
(\gen{Q}_1)^\alpha{}_b=
\varepsilon^{\alpha\gamma}\varepsilon_{bde} \PTerm{de}{\gamma},
\qquad
(\gen{S}_1)^a{}_\beta=
\varepsilon^{acd}\varepsilon_{\beta\epsilon} \PTerm{\epsilon}{cd}.
\]
Symbols $\PTerm{\cdots}{\cdots}$ with more than two indices refer to
more complex interactions.
For example, $\PTerm{\epsilon}{cd}$
replaces a sequence of two bosons $\phi^c\phi^d$ by a single fermion 
$\psi^\epsilon$.
In the model the range of interactions is bounded by the perturbative order: 
At order $g^n$ the interactions may consist of no more than $2+n$ 
spins (incoming plus outgoing), i.e.\ three in this case.

In fact, this is the leading appearance of dynamic effects within the model.
The restriction to cyclic states simplifies the 
specification of interaction symbols: In cyclic states
only the sequence of spins matters but not their overall position
along the chain. 
Thus there is no need to specify how
the final spins ($\psi^\epsilon$) are aligned with respect
to the initial spins ($\phi^c\phi^d$), 
e.g.\ left, right or centred.

These first corrections to the supercharges preserve the algebra.
The possibility of such corrections is in fact very remarkable and
related to a compatibility of the 
representation theory of cyclic chains of length $L$ and $L+1$.

\paragraph{Hamiltonian.}

The role of the Hamiltonian is somewhat special. 
It is a Cartan generator of $\alg{su}(2|3)$, 
but unlike the others its representation does receive corrections. 
Without loss of generality \cite{Beisert:2003ys}
we may assume that \eqref{eq:QSeng} holds for $\gen{H}_0$
instead of $\gen{H}(g)$
\[
\comm{\gen{H}_0}{\gen{Q}^\alpha{}_b(g)}=+\sfrac{1}{6} \gen{Q}^\alpha{}_b(g),
\qquad
\comm{\gen{H}_0}{\gen{S}^a{}_\beta(g)}=-\sfrac{1}{6} \gen{S}^a{}_\beta(g).
\]
and consequently for $\delta\gen{H}(g)=\gen{H}(g)-\gen{H}_0$
\[
\comm{\delta\gen{H}(g)}{\gen{Q}^\alpha{}_b(g)}=0,
\qquad
\comm{\delta\gen{H}(g)}{\gen{S}^a{}_\beta(g)}=0.
\]
In other words, the quantum corrections to the Hamiltonian
are invariant under the full representation of $\alg{su}(2|3)$.
In particular, the leading correction to $\gen{H}(g)$ 
must be invariant under the undeformed $\alg{su}(2|3)$ representation.
The simplest non-trivial such term is a graded permutation 
of two sites which can first appear at order $g^2$.
Together with a two-site identity operator the second order contribution reads
\[\label{eq:H2}
\gen{H}_2=
\PTerm{ab}{ab}
+\PTerm{\alpha b}{\alpha b}
+\PTerm{a\beta}{a\beta}
+\PTerm{\alpha\beta}{\alpha\beta}
-\PTerm{ba}{ab}
-\PTerm{b\alpha}{\alpha b}
-\PTerm{\beta a}{a\beta}
+\PTerm{\beta\alpha}{\alpha\beta}
.
\]
The next correction to the Hamiltonian 
appears at order $g^3$ 
\[\label{eq:H3}
\gen{H}_3=
-\varepsilon^{abc}\varepsilon_{\delta\epsilon}\PTerm{\delta\epsilon}{abc}
-\varepsilon^{\alpha\beta}\varepsilon_{cde}\PTerm{cde}{\alpha\beta}
.
\]
It is compatible with the first corrections to the supercharges 
$\gen{Q}_1$ and $\gen{S}_1$.
To some extent one can say that the Hamiltonian generally 
is shifted by two orders in $g$
with respect to the remainder of the algebra.

\paragraph{Beyond.}

The higher orders of the Hamiltonian and the algebra
have been constructed at orders $\order{g^6}$ and $\order{g^4}$,
respectively in \cite{Beisert:2003ys}. 
The concrete expressions are long and little enlightening,
but they appear to preserve integrability. 

A dynamic charge which commutes with the 
whole algebra has been derived in 
\cite{Agarwal:2005jj} at order $\order{g^1}$
providing evidence for the compatibility 
of integrability with dynamic effects.

To make integrability rigorous one could construct
the bi-local Yangian generators 
and show that they commute properly with the 
algebra and among themselves.
The Yangian generators $\gen{\hat J}$ are expected to take the generic form
\cite{Serban:2004jf,Agarwal:2004sz,Zwiebel:2006cb,Beisert:2007jv}
\[\gen{\hat J}^I{}_J\sim
\{J^I{}_K|J^K{}_J\}
-\{J^K{}_J|J^I{}_K\}
+\mbox{local},
\]
where the vertical bar stands for arbitrarily many intermediate sites
and the local terms represent a local regularisation of
the bi-local insertions. For example, the Yangian generator $\gen{\hat Q}$
corresponding to the supercharge $\gen{Q}$ reads at leading order
\[(\gen{\hat Q}_0)^\alpha{}_b\sim
\PYTerm{\alpha}{c}{c}{b}
-\PYTerm{c}{b}{\alpha}{c}
+\PYTerm{\alpha}{\gamma}{\gamma}{b}
-\PYTerm{\gamma}{b}{\alpha}{\gamma}.
\]
The first correction is expected to take the form
\[
(\gen{\hat Q}_1)^\alpha{}_b\sim
\varepsilon^{\alpha\gamma}
\varepsilon_{def}
\bigbrk{
\PYTerm{de}{\gamma}{f}{b}
-\PYTerm{f}{b}{de}{\gamma}
}
+\varepsilon^{\gamma\delta}\varepsilon_{bde}
\bigbrk{
\PYTerm{\alpha}{\gamma}{de}{\delta}
-\PYTerm{de}{\delta}{\alpha}{\gamma}
}
,
\]
where in both expressions the local regularisation terms
are very restricted and can merely be proportional to $\gen{Q}_0$
and $\gen{Q}_1$, respectively. It may be interesting to 
treat the realisation of the Yangian algebra explicitly. 
In particular, there may be complications \cite{Zwiebel:2006cb}
due to the fact that the Hamiltonian is part of the algebra itself and
because it is well-known that the Yangian is conserved
only up to boundary terms. 

\section{Undynamic Chain}

Dynamic spin chains as presented in the previous section 
have not been explored to a large extent yet. 
In this section we present an alternative formulation in 
terms of a chain with an undynamic Hamiltonian. 
The reformulation will show that
the difficulties of this particular model cannot be attributed
to the dynamic effects. 
They are rather due to the long-range nature of the interactions. 

\paragraph{Hilbert Space.}

The dynamic effects are essentially due to the 
degeneracy of quantum numbers for $\phi_{[1}\phi_2\phi_{3]}$ and
$\psi_{[1}\psi_{2]}$. The trick of freezing out the dynamic
effects consists in moving one of the bosons into the ``background''
and thus balancing the number of spins.

Let us single out one of the three bosons 
\[
\bgspin:=\phi^3
\]
and restrict Latin indices to the range $a,b=1,2$ 
for the remainder of the paper. 
We now introduce composites as the fundamental spin degrees of freedom
\[\label{eq:newspin}
\phi^a_n:=\phi^a\underbrace{\bgspin\cdots\bgspin}_{n}\,,
\qquad
\psi^\alpha_n:=\psi^\alpha\underbrace{\bgspin\cdots\bgspin}_{n}\,,
\qquad
\mdl{V}=
\bigoplus_{n=0}^\infty\,
\bigvspan{\phi^1_n,\phi^2_n\mathpunct{\big|}\psi^1_n,\psi^2_n}.
\]
Every state of the above dynamic Hilbert space can obviously be 
translated to a state of an undynamic Hilbert space
defined analogously to \eqref{eq:Hilbert}.
One simply counts the number of $\bgspin$'s following any of
the $\phi^a$ or $\psi^\alpha$ and puts as an additional index to the spin.%
\footnote{The only exceptions are the states made from $\bgspin$ alone. 
These states cannot be represented, but luckily 
they are trivial and can be ignored to a large extent.}
Note that by this redefinition
we trade in the dynamic effects 
for infinitely many spin degrees of freedom.

\paragraph{Algebra Decomposition.}

Clearly the new notation breaks the manifest $\alg{su}(3)$ symmetry
of the bosons to $\alg{su}(2)$. Together with the other $\alg{su}(2)$ 
and some of the fermionic generators the residual symmetry algebra 
reduces to $\alg{u}(2|2)$. This subalgebra is characterised by preserving
the number of spin sites and it includes the Hamiltonian. 
The remaining generators are actually still dynamic but in a controlled way: 
They either add or take away one site.

Let us decorate the residual $\alg{u}(2|2)$ generators by a tilde. 
Their embedding into $\alg{su}(2|3)$ is given by 
\[\label{eq:res}
\begin{array}[b]{rcl}
\gen{\tilde R}^a{}_b\eq\gen{R}^a{}_b+\half \delta^a_b \gen{R}^3{}_3,
\\[3pt]
\gen{\tilde L}^\alpha{}_\beta\eq\gen{L}^\alpha{}_\beta,
\end{array}
\quad
\begin{array}[b]{rcl}
\gen{\tilde Q}^\alpha{}_b\eq\gen{Q}^\alpha{}_b,
\\[3pt]
\gen{\tilde S}^a{}_\beta\eq\gen{S}^a{}_\beta,
\end{array}
\quad
\begin{array}[b]{rcl}
\gen{\tilde B}\eq\sfrac{3}{2}\gen{R}^3{}_3,
\\[3pt]
\gen{\tilde C}\eq\gen{H}-\half \gen{R}^3{}_3.
\end{array}
\]
We shall call the remaining generators dynamic and distinguish them by a hat. 
Their embedding into $\alg{su}(2|3)$ reads
\[\label{eq:dyn}
\begin{array}[b]{rcl}
\gen{\hat R}^a\eq\gen{R}^a{}_3,
\\[3pt]
\gen{\hat Q}^\alpha{}\eq\gen{Q}^\alpha{}_3,
\end{array}
\quad
\begin{array}[b]{rcl}
\gen{\hat R}_a\eq\gen{R}^3{}_a,
\\[3pt]
\gen{\hat S}_\alpha{}\eq\gen{S}^3{}_\alpha.
\end{array}
\]

The residual $\alg{u}(2|2)$ algebra is determined by the following brackets
\[\label{eq:rescomm}
\begin{array}{rcl}
\comm{\gen{\tilde B}}{\gen{\tilde Q}^\alpha{}_b}\eq
+\half\gen{\tilde Q}^\alpha{}_b,
\\[3pt]
\comm{\gen{\tilde B}}{\gen{\tilde S}^a{}_\beta}\eq
-\half\gen{\tilde S}^a{}_\beta,
\end{array}
\quad
\acomm{\gen{\tilde Q}^\alpha{}_b}{\gen{\tilde S}^c{}_\delta}=
\delta^\alpha_\delta\gen{\tilde R}^c{}_b
+\delta^c_b\gen{\tilde L}^\alpha{}_\delta
+\delta^\alpha_\delta\delta^c_b \gen{\tilde C},
\]
along with the obvious brackets of $\alg{su}(2)\times\alg{su}(2)$ generators
and trivial brackets for the central charge $\gen{\tilde C}$.
The dynamical generators form two irreducible multiplets of 
$\alg{u}(2|2)$:
$(\gen{\hat R}^a,\gen{\hat Q}^\alpha)$
and  
$(\gen{\hat R}_a,\gen{\hat S}_\alpha)$.
The non-obvious mixed brackets for the first multiplet
take the form 
\[\label{eq:mixedcomm}
\begin{array}[b]{rcl}
\comm{\gen{\tilde Q}^\alpha{}_b}{\gen{\hat R}^c}\eq
\delta^c_b\gen{\hat Q}^\alpha,
\\[3pt]
\acomm{\gen{\tilde S}^a{}_\beta}{\gen{\hat Q}^\gamma}\eq
\delta^\gamma_\beta\gen{\hat R}^a,
\end{array}
\quad
\begin{array}[b]{rcl}
\comm{\gen{\tilde B}}{\gen{\hat R}^a}\eq-\sfrac{3}{2}\gen{\hat R}^a,
\\[3pt]
\comm{\gen{\tilde B}}{\gen{\hat Q}^\alpha}\eq-\gen{\hat Q}^\alpha,
\end{array}
\quad
\begin{array}[b]{rcl}
\comm{\gen{\tilde C}}{\gen{\hat R}^a}\eq+\half\gen{\hat R}^a,
\\[3pt]
\comm{\gen{\tilde C}}{\gen{\hat Q}^\alpha}\eq+\half\gen{\hat Q}^\alpha.
\end{array}
\]
The brackets for the conjugate multiplet 
essentially follow by conjugation.
Finally, the non-trivial brackets between the 
dynamic generators yield
\[\label{eq:dynacomm}
\begin{array}[b]{rcl}
\comm{\gen{\hat R}^a}{\gen{\hat R}_b}\eq\gen{\tilde R}^a{}_b
-\delta^a_b \gen{\tilde B},
\\[3pt]
\comm{\gen{\hat R}^a}{\gen{\hat S}_\beta}\eq\gen{\tilde S}^a{}_\beta,
\end{array}
\quad
\begin{array}[b]{rcl}
\comm{\gen{\hat Q}^\alpha}{\gen{\hat R}_b}\eq\gen{\tilde Q}^\alpha{}_b,
\\[3pt]
\acomm{\gen{\hat Q}^\alpha}{\gen{\hat S}_\beta}\eq
\gen{\tilde L}^\alpha{}_\beta
+\delta^\alpha_\beta(\gen{\tilde B}+\gen{\tilde C}).
\end{array}
\]

\paragraph{Representation of the Residual Algebra.}

With the above decomposition relations it is straight-forward 
to convert the representation of the previous section 
to the new basis. The leading order $\alg{u}(2|2)$ algebra reads
\[
\begin{array}[b]{rcl}
\gen{R}^a{}_b\eq
\PTerm{a(n)}{b(n)}-\sfrac{1}{2}\delta^a_b\PTerm{c(n)}{c(n)},
\\[3pt]
\gen{L}^\alpha{}_\beta\eq
\PTerm{\alpha(n)}{\beta(n)}-\sfrac{1}{2}\delta^\alpha_\beta\PTerm{\gamma(n)}{\gamma(n)},
\end{array}\quad
\begin{array}[b]{rcl}
(\gen{Q}_0)^\alpha{}_b\eq
\PTerm{\alpha(n)}{b(n)},
\\[3pt]
(\gen{S}_0)^a{}_\beta\eq
\PTerm{a(n)}{\beta(n)},
\end{array}\quad
\begin{array}[b]{rcl}
\gen{\tilde C}_0\eq
\sfrac{1}{2}\PTerm{I(n)}{I(n)},
\\[3pt]
\gen{\tilde B}\eq
n\PTerm{I(n)}{I(n)}
-\sfrac{1}{2}\PTerm{a(n)}{a(n)}.
\end{array}
\]
Here we have extended the notation for 
interaction symbols in a hopefully evident way to 
the new states \eqref{eq:newspin},
where $n$ stands for the number of trailing $\bgspin$'s.
A repeated upper and lower index $n$ is implicitly summed over 
all integers starting from $0$. 
A capital Latin letter represents either a boson
or fermion. For example, the symbols $\PTerm{I(n)}{I(n)}$ and $n\PTerm{I(n)}{I(n)}$ count 
the length of the new chain and the number of $\bgspin$'s, respectively.

The leading correction to the supercharges reads
\<
(\gen{\tilde Q}_1)^\alpha{}_b\eq
\varepsilon^{\alpha\gamma}\varepsilon_{bd} 
\lrbrk{\PTerm{d(n+1)}{\gamma(n)}-\PTerm{I(k+1),d(n)}{I(k),\gamma(n)}},
\nln
(\gen{\tilde S}_1)^a{}_\beta\eq
\varepsilon^{ac}\varepsilon_{\beta\delta} 
\lrbrk{\PTerm{\delta(n)}{c(n+1)}-\PTerm{I(k),\delta(n)}{I(k+1),c(n)}}.
\>
While in \eqref{eq:QS1} all interactions were one-to-two or two-to-one site,
here we get one-to-one site or two-to-two site operators. In the case
of the two-to-two site contributions the second site is merely needed 
to account for the change of leading $\bgspin$'s which cannot
be represented otherwise.  

A careful conversion of the leading interacting Hamiltonian \eqref{eq:H2}
yields the new representation
\<
\gen{\tilde C}_2
\eq
\PTerm{I(k),J(n+1)}{I(k),J(n+1)}
-\PTerm{I(k+1),J(n)}{I(k),J(n+1)}
-\PTerm{I(k),J(n+1)}{I(k+1),J(n)}
+\PTerm{I(k+1),J(n)}{I(k+1),J(n)}
\nl
+\PTerm{I(0),J(n)}{I(0),J(n)}
-\PTerm{a(0),b(n)}{b(0),a(n)}
-\PTerm{\alpha(0),b(n)}{b(0),\alpha(n)}
-\PTerm{b(0),\alpha(n)}{\alpha(0),b(n)}
+\PTerm{\beta(0),\alpha(n)}{\alpha(0),\beta(n)}.
\>
Gladly, this is still a nearest-neighbour spin chain Hamiltonian.
Note that the terms on the two above lines have a somewhat different meaning:
The terms on the first row represent propagation terms of the magnons 
along the original chain, while the terms on the second row represent
spin interactions of two adjacent magnons.
The first correction to the interacting Hamiltonian \eqref{eq:H3}
marks the leading appearance of dynamic effects. 
In the new basis, however, the length remains fixed
\<
\gen{\tilde C}_3\eq
\varepsilon_{cd}\varepsilon^{\alpha\beta}
\lrbrk{
-\PTerm{c(0),d(n+1)}{\alpha(0),\beta(n)}
+\PTerm{c(1),d(n)}{\alpha(0),\beta(n)}
-\PTerm{I(k+1),c(0),d(n)}{I(k),\alpha(0),\beta(n)}
}
\nl
+\varepsilon^{cd}
\varepsilon_{\alpha\beta}
\lrbrk{
-\PTerm{\alpha(0),\beta(n)}{c(0),d(n+1)}
+\PTerm{\alpha(0),\beta(n)}{c(1),d(n)}
-\PTerm{I(k),\alpha(0),\beta(n)}{I(k+1),c(0),d(n)}
}.
\>

\paragraph{Representation of Dynamic Generators.}

Note that $\gen{C}_0$ measures half the length of the undynamic chain
and thus the two brackets in \eqref{eq:mixedcomm} 
imply that the generators
$\gen{\hat R}^a$ and $\gen{\hat Q}^\alpha$ add one site
while 
$\gen{\hat R}_a$ and $\gen{\hat S}_\alpha$ remove one site.
The leading-order representation takes the form
\[
\begin{array}{rcl}
\gen{\hat R}^{a}\eq
\PTerm{I(k),a(n-1-k)}{I(n)},
\\[3pt]
(\gen{\hat Q}_0)^{\alpha}\eq
\PTerm{I(k),\alpha(n-1-k)}{I(n)},
\end{array}\quad
\begin{array}{rcl}
\gen{\hat R}_{a}\eq
\PTerm{I(n)}{I(k),a(n-1-k)},
\\[3pt]
(\gen{\hat S}_0)_{\alpha}\eq
\PTerm{I(n)}{I(k),\alpha(n-1-k)},
\end{array}
\]
which changes the length by one unit, because they
replace a background spin $\bgspin$ by something else or
vice versa.

Despite the length fluctuation, these generators close onto
the one-to-one generators of the residual $\alg{u}(2|2)$ representation. 
For example the non-manifest $\alg{su}(3)$ brackets can
be performed easily
\<
\comm{\gen{\hat R}^{a}}{\gen{\hat R}_{b}}
\eq
\sum_{m=0}^\infty
\sum_{n=0}^\infty
\sum_{k=0}^{n-1}
\sum_{j=0}^{m-1}
\lrcomm{\PTerm{I(k),a(n-1-k)}{I(n)}}{\PTerm{J(m)}{J(j),b(m-1-j)}}
\nln\eq
\sum_{k=0}^{\infty}
\sum_{n=0}^\infty
\PTerm{I(k),a(n)}{I(k),b(n)}
-\sum_{n=0}^\infty
\delta^a_b n\PTerm{I(n)}{I(n)}
=\gen{\tilde R}^{a}{}_{b}
-\delta^a_b\gen{\tilde B},
\>
as it should according to \eqref{eq:dynacomm}.

The first correction to the dynamic supercharges reads
\[
(\gen{\hat Q}_1)^\alpha{}=
\varepsilon^{\alpha\beta}\varepsilon_{cd} \PTerm{c(0),d(n)}{\beta(n)},
\qquad
(\gen{\hat S}_1){}_\alpha=
\varepsilon^{cd}\varepsilon_{\alpha\beta} \PTerm{\beta(n)}{c(0),d(n)}.
\]
Actually, it is not necessary to specify  
either of the pairs $\gen{\hat Q},\gen{\hat S}$ 
or $\gen{\tilde Q},\gen{\tilde S}$ explicitly 
because according to \eqref{eq:mixedcomm,eq:dynacomm}
one pair can simply be obtained from the other by 
commutation with the exact generators $\gen{\hat R}$.

Dynamic (super)symmetries which relate 
conventional nearest-neighbour spin chain models 
at lengths differing by one unit are not unheard of: 
In particular they have appeared 
in various sectors of AdS/CFT 
\cite{Beisert:2004ry,Beisert:2005fw,Zwiebel:2005er,Beisert:2007sk}.
They also exist for the XXX$_1$ chain \cite{Schoutens:2008aa}, 
the XXZ$_{1/2}$ chain with $q=e^{\pm 2\pi i/3}$
\cite{Fendley:2003je,Fendley:2006mj} 
(or more generally XXZ$_s$ with $q=e^{n\pi i/(1\pm s)}$),
and a more exotic model \cite{Santachiara:2005aa}.
They all share the feature that Bethe roots
at rapidity $0$ induce the symmetry and that the
symmetry can only exist for cyclic closed chains
or for open chains.

\section{Comments}

In this final section I would like to comment 
on the reformulation performed in the previous 
section and on the possibility of
extending such a reformulation to the
whole AdS/CFT spin chain with $\alg{psu}(2,2|4)$ symmetry. 

\paragraph{Algebraic Formulation.}

It is fair to say that the picture presented in the previous section 
does not constitute an improvement of the situation per se.
For example, the construction of \cite{Beisert:2003ys}
would not simplify in the new basis. In fact it would be somewhat worse,
because the range of the interactions changes drastically between the pictures:
The perturbative construction is expected to follow the range of the original 
spin chain, while the range in the new basis 
represents the number of magnon excitations involved in the interaction. 
Moreover, the manifest $\alg{su}(3)$ symmetry
reduces to merely $\alg{su}(2)\times \alg{u}(1)$. 
Finally, there is an unaesthetic asymmetry 
between leading and trailing background spins $\bgspin$.
Nevertheless, there is a one-to-one map of interactions, and thus 
essentially nothing is lost by the change of picture.

The potential advantages of the new basis are of a more formal nature. 
There is some hope that the absence of dynamic effects for a large part 
of the algebra will make the model more accessible to conventional 
algebraic methods such as a Hopf algebra treatment. For example,
to represent the action of symmetry generators through a coproduct
appears reasonable only if the representation is undynamic.
However, the problems introduced by 
long-range interactions certainly remain to be overcome. 
The remaining dynamic symmetry generators
change the length by exactly one unit, 
which is much better than the arbitrariness in the original picture.
In fact the various symmetry enhancements discussed in 
\cite{Fendley:2003je,Santachiara:2005aa,Fendley:2006mj}
and \cite{Beisert:2004ry,Beisert:2005fw,Beisert:2007sk}
are of or can be brought to this form and they 
call for a more general story yet to be understood.

\paragraph{Excitation Picture.}

The picture advertised here closely resembles 
what one obtains by performing the coordinate Bethe ansatz
in \cite{Beisert:2005tm}.%
\footnote{A similar picture 
also underlies the NLIE approach, see e.g.\ \cite{Bombardelli:2007ed,Freyhult:2007pz}.}
Namely, the spin $\bgspin$ is treated 
as a background spin and the magnons
become sites of the reduced chain 
with $2|2$ spin orientations per site.
The only difference is that magnons carry a definite momentum while 
here the spin orientation also specifies 
the distance between two adjacent excitations.%
\footnote{\label{fn:1}It might also be worthwhile to investigate an absolute
(instead of relative) position space picture, 
where, however, length fluctuations may become difficult to handle.}
In that sense, these two pictures are essentially related 
by Fourier transformation.

In fact, the residual $\alg{u}(2|2)$ algebra acting on the new basis 
coincides with the $\alg{su}(2|2)$ algebra of the 
coordinate Bethe ansatz.
In this context the difference between the pictures is that here UV effects, 
i.e.\ what happens when two magnons come close (in the original model), 
can be honestly represented.
This may be crucial for understanding finite-size effects.
In the asymptotic coordinate Bethe ansatz such effects are largely 
ignored and collectively accounted for by the S-matrix.
Conversely, here it is not possible to represent gauge transformations
in a consistent manner. In the coordinate Bethe ansatz the gauge transformations 
alias the central extensions were crucial for success of the construction.
Representing the $\alg{su}(2|2)$ algebra within the coordinate Bethe ansatz
is particularly simple because one only has to understand the 
single-magnon representation and how to assemble multi-magnon
representations from that. The latter is achieved by 
a coproduct \cite{Gomez:2006va,Plefka:2006ze}
within the Hopf algebra framework.
One might actually do the same here, at least to some approximation:
Namely, find a representation of $\alg{u}(2|2)$ on the infinite-dimensional 
spin module. A similar proposal has appeared recently for the closely
related exceptional superalgebra $\alg{d}(2,1;\alpha)$ in \cite{Matsumoto:2008ww}.

\paragraph{Complete AdS/CFT Spin Chain.}

It would be desirable to represent the complete
AdS/CFT spin chain with $\alg{psu}(2,2|4)$ symmetry.
However, the generalisation is not straight-forward: 
The spin module $\mdl{V}$ contains not only the background spin and
single excitations, but also multiple excitations. 
The decomposition of $\mdl{V}$ in terms of the subalgebras
$\alg{psu}(2|2)\times\alg{psu}(2|2)$ reads
\cite{Beisert:2006qh}
\[
\mdl{V}=\bigoplus_{n=0}^\infty \mdl{V}^{}_n\otimes\mdl{V}'_n,
\qquad
\mdl{V}^{}_n=\lreval{\lrbrk{\mdl{V}^{}_1}^{\otimes n}}\indup{antisym}.
\]
Here $\mdl{V}_0$ is the trivial module spanned by the background spin $\bgspin$
and $\mdl{V}_n$ is the $n$-fold graded anti-symmetric 
tensor product of $\mdl{V}_1=\vspan{\phi^1,\phi^2\mathpunct{|}\psi^1,\psi^2}$.
The $\mdl{V}'_n$ denote the corresponding modules of the second $\alg{psu}(2|2)$.
There are now two ways in which one could attempt to proceed:

As before one could dress each of the components
$\mdl{V}^{}_n\otimes\mdl{V}'_n$ for $n\neq 0$ 
by an arbitrary number of background spins $\bgspin$.
However this would not freeze the spin chain
because only the overall number of excitations $n$ is conserved.
For example, two single excitations ($n=1$) can be mapped to one
double excitation ($n=2$). 

Instead one should work only with $\mdl{V}^{}_1\otimes\mdl{V}'_1$
trailed by arbitrarily many background spins $\bgspin$. 
The higher excitations would be represented 
by gluing together single excitations. 
For example the double excitation $\bar\bgspin$ can be 
thought of as being composed from $\phi^b\otimes\phi^{\dot a}$ 
and $\phi^d\otimes\phi^{\dot c}$:
\[
\bar\bgspin_n
\to
\varepsilon_{bd}\varepsilon_{\dot a\dot c}\,
(\phi^b\otimes\phi^{\dot a})_{-1}
(\phi^d\otimes\phi^{\dot c})_n.
\]
Here the number ``$-1$'' of trailing $\bgspin$'s is meant
to indicate that the two consecutive single excitations 
reside on a single site 
(i.e.\ $-1$ sites in between)
and thus form a double excitation. 
The problem with this representation
is the graded anti-symmetrisation implicit for multiple excitations. 
Consequently the Hilbert space $\mdl{H}_L$ of the model contains
additional unphysical states.%
\footnote{The same problem arises in an absolute position space picture, 
cf.\ footnote \ref{fn:1} on page \pageref{fn:1},
when the excitations are not well-ordered.} 
Therefore one has to ensure that the Hamiltonian 
and the symmetry generators do not map 
physical states to unphysical states.
One could project out unphysical states from the 
Hilbert space from the start. 
This would lead to potential problems with the 
definition of interactions (they have to be compatible with the projection).
Alternatively one could adjoin the Hamiltonian
with a projection onto physical states. 
Unfortunately the latter are defined in a long-ranged fashion
(an arbitrary number of adjacent spins has to be symmetrised). 
This apparently makes even the leading-order Hamiltonian long-ranged.

\paragraph{Conclusions.}

In conclusion, I have presented a reformulation 
of the $\alg{su}(2|3)$ dynamic spin chain constructed
in \cite{Beisert:2003ys} where the dynamic effects are 
frozen out for a $\alg{u}(2|2)$ subalgebra
including the Hamiltonian.
The other generators remain dynamic, but they 
merely change the length by precisely one unit
as in \cite{Fendley:2003je,Fendley:2006mj,Santachiara:2005aa,Beisert:2004ry,Beisert:2005fw,Zwiebel:2005er,Beisert:2007sk}.
The reformulation is intended to make the chain more accessible
to a conventional algebraic treatment;
it is merely the first step. 

The change of picture works nicely for
$\alg{su}(2|3)$ where only single excitations
of the ferromagnetic vacuum exist. 
A similar treatment of the complete AdS/CFT spin chain 
with $\alg{psu}(2,2|4)$ symmetry
and infinite-dimensional spin representations
requires further insight due
to the existence of multiple coincident excitations.
However, if the proposed undynamic reformulation
leads to a better understanding of the $\alg{su}(2|3)$ model 
then there may well be a way to generalise those
results to $\alg{psu}(2,2|4)$.

\paragraph{Acknowledgements.}

I would like to thank the Galileo Galilei Institute
(workshop ``Non-Per\-tur\-ba\-tive Methods
in Strongly Coupled Gauge Theories''),
Institut des Hautes \'Etudes Scientifiques
and in particular Kyoto University 
(conference ``30 Years of Mathematical Methods in High Energy Physics'')
for hospitality while part of this work was performed.

\bibliography{undynamic}
\bibliographystyle{nb}

\end{document}